\documentclass[prb, aps, floatfix, twocolumn]{revtex4} 
\usepackage{graphicx,bm}
\usepackage{subfigure} 
\usepackage{amssymb} 
\usepackage{amsmath} 
\usepackage{verbatim} 
\usepackage{my_symbols} 
\usepackage{amsfonts} 
\usepackage{natbib}

\newcommand{\Heff}{\ensuremath{{\bf H}_{\rm eff}}}
\newcommand{\mmx}{\ensuremath{m_x}}
\newcommand{\mmy}{\ensuremath{m_y}}

\newcommand{\dmmx}{\ensuremath{\dot{m}_x}}
\newcommand{\dmmy}{\ensuremath{\dot{m}_y}}

\newcommand{\bmu}{\bm {\mu}} \newcommand{\II}{{\bf I}}

\begin{document}

\title{Charge pumping and the colored thermal voltage noise in spin valves}

\author{Jiang Xiao,$^1$ Gerrit E.W. Bauer,$^1$ Sadamichi Maekawa,$^{2,3}$ and Arne
Brataas$^4$} \affiliation{$^1$Kavli Institute of NanoScience, Delft University of Technology,
2628 CJ Delft, The Netherlands \\ $^2$Institute for Materials Research, Tohoku University,
Sendai, Japan\\ $^3$CREST, Japan Science and Technology Agency, Tokyo 100-0075, Japan\\
$^4$Department of Physics, Norwegian University of Science and Technology, NO-7491 Trondheim,
Norway}

\begin{abstract}

Spin pumping by a moving magnetization gives rise to an electric voltage over a spin valve.
Thermal fluctuations of the magnetization manifest themselves as increased thermal voltage
noise with absorption lines at the ferromagnetic resonance frequency and/or zero frequency. The
effect depends on the magnetization configuration and can be of the same order of magnitude as
the Johnson-Nyquist thermal noise. Measuring colored voltage noise is an alternative to
ferromagnetic resonance experiments for nano-scale ferromagnetic circuits.

\end{abstract} 
\maketitle


\section{Introduction} \label{sec:intro}

A spin valve consists of a thin non-magnetic metallic layer (NM) sandwiched by two
ferromagnetic layers (FM) with variable magnetization direction. One of the FM layers is
usually thick and its magnetization is fixed, while the other is thin and its magnetization
direction is free to move. Spin valves have a wide range of interesting static and dynamic
properties, \cite{slonczewski_current-driven_1996, berger_emission_1996,
kiselev_microwave_2003, katine_current-driven_2000, stiles_spin-transfer_2006,
urazhdin_current-driven_2003, rippard_direct-current_2004, fert_magnetization_2004,
sun_spin-current_2000, huertas_hernando_conductance_2000, waintal_role_2000,
stiles_anatomy_2002, li_magnetization_2003} many of which are related to the current-induced
spin-transfer torque, \cite{slonczewski_current-driven_1996, berger_emission_1996} which can
excite magnetization dynamics (and reversal). Inversely, magnetization dynamics generates a
current flow or a voltage output. Berger first discussed the induced voltage in an
FM\big|NM\big|FM structure by magnetization dynamics. \cite{berger_generation_1999} He posited
that a voltage of order ${\hbar}{\omega}/e$ can be generated when the magnetization of one ferromagnet
precesses at frequency ${\omega}$. Similar dynamically induced voltages have been studied
theoretically \cite{wang_voltage_2006} and observed \cite{costache_electrical_2006} in simple
FM\big|NM junctions and in magnetic tunnel junctions (MTJs). \cite{xiao_charge_2008} In spin
valves and MTJs, voltage induced by the magnetization dynamics can be understood as a two-step
process: i) the moving magnetization of the free layer generates a spin current; ii) the static
magnetization of the fixed layer filters the ``pumped'' spin current and converts it into a
charge current or, in an open circuit, a voltage output. The electrical voltage induced by
moving domain walls can be explained analogously. \cite{barnes_generalization_2007,
saslow_spin_2007, duine_spin_2008, tserkovnyak_electron_2008, yang_universal_2009} In the first
part of the present paper, we derive a simple formula for the charge pumping voltage in a spin
valve by circuit theory in which magnetization dynamics is taken into account. We find that the
magnitude of the voltage is governed by the spin-transfer torque in the same structure. We
therefore propose to measure the angular dependence of the spin-transfer torque (or torkance,
{\it i.e.} the torque divided by the voltage bias) by the angular dependence of the charge
pumping voltage.

The charge pumping voltage consists of a DC and an AC component, even when induced by a steady
magnetization dynamics such as ferromagnetic resonance (FMR). The concept can be extended to
the case of thermally activated, {\it i.e.} fluctuating, magnetization dynamics, which is an
extra source of thermal voltage noise that only appears in magnetic structures. Johnson
\cite{johnson_thermal_1928} and Nyquist \cite{Nyquist_thermal_1928} showed that in non-magnetic
conductors the voltage noise is associated with thermal agitation of charge carriers (driven by
fluctuating electromagnetic modes). The power spectrum of this noise is white and proportional
to the temperature $T$ and resistance $R$: $S_{JN}({\omega}) = 4\kB TR$ up to frequencies of $\kB
T/{\hbar}{\sim}10^4$ GHz at room temperature. \cite{johnson_thermal_1928, Nyquist_thermal_1928} In
magnetic structures such as spin valves thermal fluctuations of the magnetization direction
have to be considered. \cite{brown_thermal_1963} Some consequences of thermal fluctuation in
spin valves, such as noise-facilitated magnetization switching \cite{koch_thermally_2000,
wetzels_efficient_2006, wang_current_2008} and resistance fluctuations,
\cite{foros_magnetization_2005, foros_resistance_2007} have been studied before. In the so
called thermal ferromagnetic resonance, frequencies are studied by means of resistance
fluctuations without applied magnetic fields. \cite{deac_bias-driven_2008} Foros {\it et al.}
\cite{foros_resistance_2007} have shown that the time-averaged auto-correlator of the
resistance fluctuations is significantly affected by dynamical exchange coupling between the
magnetic layers. 

In the second part of this paper, we show that a magnetization fluctuation-related voltage
noise can be of the same order of magnitude as the conventional thermal noise in non-magnetic
conductors. This noise is not ``white'' but displays spectral features related to the
ferromagnetic resonance (FMR). The noise spectrum therefore contains information comparable to
that obtained by FMR. For nano-scale ferromagnetic circuits the noise measurements might be
easier to perform than conventional FMR experiments. Compared to the resistance noise
measurement, the pumping voltage noise measurement is non-intrusive because it does not require
application of current, which may disturb the system.

The paper is organized as follows: Section II presents a general theoretical frame work that
combines the magnetoelectric circuit theory and the Landau-Lifshitz-Gilbert (LLG) equation. In
Section III, we derive a formula for the charge pumping voltage in spin valves. In Section IV,
by using the magnetic susceptibility function, we calculate the voltage noise spectrum due to
magnetization fluctuations for two different magnetic configurations. Section V contains some
general remarks on the calculation. In Appendix A we calculate the angular-dependence of the
magnetic susceptibility for a spin valve, and Appendix B presents an alternative calculation of
the magnetization-related thermal noise by computing the frequency dependent impedance of a
spin valve.

\section{Circuit theory with dynamics} \label{sec:ct}

\begin{figure}[b]
	\includegraphics[width=0.7\columnwidth]{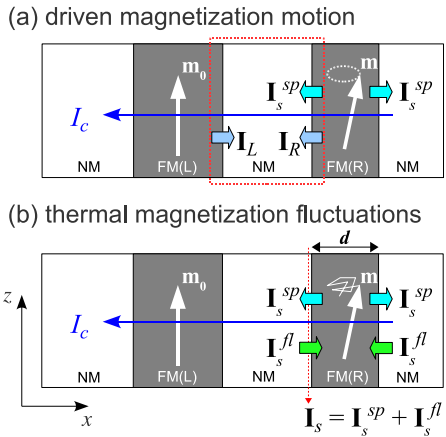}
	\caption{(Color online) Spin and charge currents in spin valves. (a) For the steady state
	case studied in Section \ref{sec:ct} and \ref{sec:cpc}. (b) For the thermal magnetization
	fluctuations studied in Section \ref{sec:spec}.}
	\label{fig:spinvalve}
\end{figure}

\Figure{fig:spinvalve}(a) shows a spin valve structure under consideration. The magnetization
in the left FM with direction $\mm_0$ is assumed to be static and $\mm$, the one of the right
FM, to be free, which can be realized by making the right layer much thinner than the left one.
For electron transport, we assume for simplicity that the spin valve is symmetric. Such an
assumption may be invoked when both FM layers are of the same material and thicker than the
spin flip diffusion length. In that regime, the resistances of the bulk ferromagnet over the
spin-flip diffusion length are in series with the interface resistances, whereas the remoter
parts of the ferromagnets are magnetically inert series resistances. The regime in which the
layers become thinner than the spin flip diffusion length has been treated by Kovalev {\it et.
al.} \cite{kovalev_current-driven_2007} In this and the next Section, we focus on the
spin-active region in the spin valve, which includes the NM spacer and small part (of the order
of the spin-flip diffusion length) of the FM layers as indicated by the dashed box in
\Figure{fig:spinvalve}(a).

When $\mm$ depends on time, a spin current is pumped into the metallic spacer layer through the
F\big|N interface. The magnitude and polarization of the spin pumping current reads:
\cite{tserkovnyak_enhanced_2002}
\begin{equation}
	{\bf I}_s^{sp} = {{\hbar}\over 4{\pi}}\smlb{g_r~\mm{\times}{d\mm\over dt} + g_i{d\mm\over dt}},
\label{eqn:Isp}
\end{equation}
where $g_r$ and $g_i$ are the real and imaginary part of the dimensionless transverse
spin-mixing conductance. \cite{tserkovnyak_enhanced_2002} The first term in $\II_s^{sp}$
corresponds to a loss of angular momentum of the free layer magnetization to the adjacent NM
layers, thus providing an extra damping torque. \cite{tserkovnyak_enhanced_2002} When the
adjacent normal metal is an ideal spin sink, the spin pumping current loss can be represented
by a Gilbert damping coefficient (see below), and each interface contributes to the damping
constant by ${\alpha}' = ({\gamma}{\hbar}/4{\pi} M_{tot}) g_r$, introducing the gyromagnetic ratio ${\gamma}$ and the
total magnetization of the right FM film $M_{tot}$. The imaginary part $g_i$ effectively
modifies the gyromagnetic ratio for the magnetization under consideration.

In order to use magnetoelectronic circuit theory, the structure has first to be decomposed into
nodes (for bulk) and contacts (for interfaces). For each node we may define a charge chemical
potential and a (vector) spin chemical potential: let ${\mu}_{L, R, N}$ and $\bmu_{L, R, N}$ be
the charge and spin chemical potentials in left, right FM leads, and the NM spacer. In the
ferromagnet, we may assume that the spin accumulations are aligned with the magnetization, {\it
i.e.} $\bmu_L = {\mu}_L^s \mm_0$ and $\bmu_R = {\mu}_R^s \mm$. The charge current $I_c$ and the spin
current $\II_L$ through the left interface connecting the left FM and the spacer layer are
given by: \cite{brataas_finite-element_2000, brataas_spin-transport_2001}
\begin{subequations} 
\label{eqn:ctL}
\begin{align}
	I_c &= {eg\over 2h}\midb{2({\mu}_L-{\mu}_N)+p({\mu}_L^s-\bmu_N{\cdot}\mm_0)}, 
	\label{eqn:IcL}\\
	\II_L &= -{eg\over 2h}\midb{2p({\mu}_L-{\mu}_N)+({\mu}_L^s-\bmu_N{\cdot}\mm_0)}\mm_0 \nn 
	&+{e\over h}g_r\mm_0{\times}\bmu_N{\times}\mm_0+{e\over h}g_i\bmu_N{\times}\mm_0, 
	\label{eqn:IsL}
\end{align}
\end{subequations}
where $g = g_\up + g_\dn$ is the total conductance and $p = (g_\up-g_\dn)/g$ is the
polarization of the F\big|N interface and the (longitudinal) active regions of the FMs.
\cite{kovalev_current-driven_2007} Similarly for the right FM lead:
\begin{subequations} 
\label{eqn:ctR}
\begin{align}
	I_c &= -{eg\over 2h}\midb{2({\mu}_R-{\mu}_N)+p({\mu}_R^s-\bmu_N{\cdot}\mm)}, \label{eqn:IcR} \\
	\II_R &=
	-{eg\over 2h}\midb{2p({\mu}_R-{\mu}_N)+({\mu}_R^s-\bmu_N{\cdot}\mm)}\mm \nn 
	&+{e\over h}g_r\mm{\times}\bmu_N{\times}\mm +{e\over h}g_i\bmu_N{\times}\mm + {2e\over{\hbar}}{\bf I}_s^{sp},
	\label{eqn:IsR}
\end{align}
\end{subequations}
where the spin current $\II_R$ at the right interface is modified due to the additional spin
pumping current $\II_s^{sp}$ emitted by the moving magnetization $\mm$. Additionally, the spin
current conservation in the presence of the spin flips in the NM spacer requires:
\begin{equation}
	\label{eqn:gsf}
	\II_L + \II_R = {e\over h}{h\over D{\tau}_{sf}}\bmu_N {\equiv} {e\over h}g_{sf}\bmu_N,
\end{equation}
where $D$ the energy density of states at the Fermi energy and ${\tau}_{sf}$ the spin flip
relaxation time in the NM.

The spin current $\II_R$ entering the free layer exerts a spin-transfer torque on $\mm$, which
is equal to its transverse component absorbed at the interface: \cite{waintal_role_2000,
stiles_anatomy_2002}
\begin{equation}
	\label{eqn:Nst}
	{\bf N}_{\rm st}= {{\hbar}\over 2e}\midb{\II_R - (\II_R{\cdot}\mm)\mm)}. 
\end{equation}
The LLG equation is therefore modified as:
\begin{equation}
	{d\mm\over dt} = -{\gamma}~\mm{\times}\Heff + {\alpha}~\mm{\times}{d\mm\over dt} + {{\gamma}\over M_{tot}}{\bf N}_{\rm st},
	\label{eqn:llg}
\end{equation}
where $\Heff$ is the total effective magnetic field acting on $\mm$, and ${\alpha} = {\alpha}_0 + 2{\alpha}'$ is
the total magnetic damping including both the bulk damping and the spin pumping enhanced
damping from both interfaces.

The set of equations \Eqss{eqn:Isp}{eqn:llg} describes the charge/spin transport and
magnetization dynamics in the metallic magnetic heterostructures. In many cases, the transport
equations \Eqss{eqn:Isp}{eqn:Nst} and the dynamical LLG equation \Eq{eqn:llg} can be solved
separately by ignoring the spin pumping contribution $\II_s^{sp}$ in $\II_R$, in which case the
transport only depends on the instantaneous $\mm$ but not on $\dmm$. However, for thin magnetic
layers the spin pumping modification cannot be neglected. It has possibly important
consequences, such as a voltage induced by magnetization dynamics, anisotropic magnetic damping
and susceptibility tensor, and colored thermal noise, as will become clear from the discussion
below.


At first, let us calculate the static ($\dmm = 0$) magneto conductance of a spin valve. When a
bias voltage $V = ({\mu}_L - {\mu}_R)/e$ is applied, we can calculate the charge current $I = I_L =
I_R$ from \Eqss{eqn:ctL}{eqn:gsf}, hence the magneto-conductance $G = I/V$. By setting ${\mu}_N =
0$ in the spacer and assuming strong spin flips in the ferromagnets (${\mu}_L^s = {\mu}_R^s = 0$), we
find (with $\mm{\cdot}\mm_0 = \cos{\theta}$)
\begin{equation}
	G({\theta}) =
	{G_0g\over 4}\midb{1 - 4p{\eta}({\theta})\sin^2{{\theta}\over2}},
	\label{eqn:GMR}
\end{equation}
where $G_0 = 2e^2/h$ is the conductance quantum and
\begin{equation}
	{\eta}({\theta}) = {pg/4 \over
	g \sin^2{{\theta}\over2} +2\td{g}_r \cos^2{{\theta}\over 2}+ g_{sf}}
\end{equation}
is the angular dependent spin current polarization with $\td{g}_r {\equiv}g_r +
2g_i^2/(2g_r+g_{sf})$. The $G({\theta})$ above agrees with Eq. (160) in
Ref.~\onlinecite{brataas_non-collinear_2006}.

\section{Charge pumping in spin valves} \label{sec:cpc}

When a voltage difference ${\Delta}V$ is applied over a spin valve which does not excite
magnetization dynamics ($\dmm = 0$), \Eqss{eqn:ctL}{eqn:Nst} lead to the spin-transfer torque:
\cite{slonczewski_current-driven_1996}
\begin{equation}
	\label{eqn:Nst2}
	{\bf N}_{st}({\theta}) = {\Delta}V\midb{{\tau}_{ip}({\theta}) \mm{\times}\mm_0 + {\tau}_{op}({\theta})\mm_0}{\times}\mm,
\end{equation}
where ${\tau}_{ip}$ and ${\tau}_{op}$ are the so-called (angular-dependent) torkances
\cite{slonczewski_theory_2007} for the in-plane (Slonczewski's) component and out-of-plane
(effective field) component:
\begin{equation}
	{\tau}_{ip}({\theta}) = {e{\eta}({\theta})\over 2{\pi}}\td{g}_r \qand
	{\tau}_{op}({\theta}) = {e{\eta}({\theta})\over 2{\pi}}{g_ig_{sf}\over 2g_r+g_{sf}}.
\end{equation}
When the bias polarity is chosen such that the in-plane torque in \Eq{eqn:Nst2} works against
the magnetic damping, the current flow can excite magnetization dynamics, otherwise they are
suppressed.

Inversely, magnetization dynamics can induce a current flow by the spin-pumping: a moving
magnetization ($\mm$) pumps a spin current (with zero charge current) into adjacent contacts,
and the pumped spin current is converted into a charge current $I_P$ (or pumping voltage $V_P$)
by a static ferromagnetic filter ($\mm_0$). \cite{xiao_charge_2008} In the following, we use
the circuit theory described in Section \ref{sec:ct} to derive a simple expression for the
charge pumping voltage (current) induced by FMR in a spin valve. We shall study two different
cases, i) when the spin valve is open, so no current flow is allowed ($I_c = 0$), and a pumping
voltage $V_P$ is built up; ii) when the spin valve is closed, {\it i.e.} the two ends of the
spin valve are short-circuited, so no voltage difference is allowed at the two ends (${\mu}_L =
{\mu}_R$), and a pumping current $I_P$ flows.

i) open circuit - For an open circuit, the charge current vanishes: $I_c = 0$. By solving
\Eqss{eqn:Isp}{eqn:gsf}, we find an electric voltage $V_P = ({\mu}_L-{\mu}_R)/e$ due to the spin
pumping current $\II_s^{sp}$:
\begin{equation}
	V_P ({\theta})
	= R({\theta}) \midb{{\tau}_{ip}({\theta})\mm{\times}\dmm + {\tau}_{op}({\theta})\dmm}{\cdot}\mm_0,
	\label{eqn:VP}
\end{equation}
with the magneto-resistance $R({\theta}) = 1/G({\theta})$. Both the spin-transfer torque in \Eq{eqn:Nst2}
and the charge pumping voltage in \Eq{eqn:VP} are governed by the torkances. \Eq{eqn:VP}
confirms the two-step process for the charge pumping: 1) spin current pumped by $\dmm$, 2)
charge current generated by projecting on $\mm_0$. Note that \Eq{eqn:VP} entails all multiple
scattering in the spacer.

In \Eq{eqn:VP}, the charge pumping voltage is related to the torkances, which also govern the
spin-transfer torque. Currently, the latter can be accessed only indirectly by its effect on
the current induced magnetization dynamics for MTJs. \cite{kubota_quantitative_2008,
sankey_measurement_2008} \Eq{eqn:VP} can be employed to measure the spin-transfer torque or
torkances in spin valves via FMR induced voltages when the magneto-resistance $R({\theta})$ is
obtained alongside as done by Urazhdin {\it et al}. \cite{urazhdin_switching_2004}

ii) closed circuit - When the spin valve is short-circuited, ${\mu}_L = {\mu}_R$,
\Eqss{eqn:Isp}{eqn:gsf} yield a pumping current
\begin{equation}
	I_P({\theta}) = \midb{{\tau}_{ip}({\theta})\mm{\times}\dmm + {\tau}_{op}({\theta})\dmm}{\cdot}\mm_0.
	\label{eqn:IP}
\end{equation}
For comparison, in the presence of an applied current current $I$,
\begin{equation}
	{\bf N}_{st}({\theta}) = R({\theta})I\midb{{\tau}_{ip}({\theta}) \mm{\times}\mm_0 + {\tau}_{op}({\theta})\mm_0}{\times}\mm.
\label{eqn:NstI}
\end{equation}

Usually, there is a passive series resistance in addition to the magneto-resistance $R({\theta})$.
The charge pumping voltage for the open circuit is insensitive to such a passive resistance
(for an ideal voltage meter).

In transition-metal ferromagnets, $g_i\le 0.1g_r$, \cite{brataas_non-collinear_2006} thus from
now on we neglect the imaginary part of the mixing conductance, {\it i.e.} $g_i = 0$, ${\tau}_{op}
= 0$, and ${\tau}_{ip} = e{\eta}({\theta})g_r/2{\pi}$.

\section{Magnetization related voltage noise in spin valves} \label{sec:spec}

According to the Fluctuation-Dissipation theorem (FDT), the electrical voltage fluctuations
across a non-magnetic conductor is associated with the electron linear momentum dissipation by
the electrical resistance, which causes Joule heating. In ferromagnets, there are also
magnetization fluctuations associated with the angular momentum dissipation or magnetic
damping. In magnetic heterostructures such as spin valves, the two fluctuations (electric and
magnetic) are coupled by the dynamical exchange of spin currents. Electronic noise increases
the magnetic fluctuations via the spin-transfer effect. Inversely, the magnetization noise
increases electronic fluctuations via spin/charge pumping.

We discussed in the previous section the pumping voltage (current) induced by an arbitrary
motion of magnetization. Here the formalism is applied to the stochastic magnetization motion
at thermal equilibrium: the thermal fluctuations of magnetization induce a pumping voltage
(current), which on filtering by the static layer becomes a noisy voltage signal. In this
section, we discuss this magnetization fluctuation-induced voltage noise $V_M(t)$ in a spin
valve at thermal equilibrium, the power spectrum of which is the Fourier transform of its time
correlation function:
\begin{equation}
	S_M({\omega}) = 2{\int}\avg{V_M(0)V_M(t)}e^{-i{\omega}t}dt.
	\label{eqn:Somega}
\end{equation}


As shown by Johnson and Nyquist, \cite{johnson_thermal_1928, Nyquist_thermal_1928} the
Fluctuation-Dissipation theorem (FDT) relates the noise power spectrum $S({\omega})$ to the real part
of the impedance $Z({\omega})$ which characterizes the dissipation:
\begin{equation}
	S({\omega}) = 4k_BT~\rem{Z({\omega})}.
\label{eqn:JN}
\end{equation}
We may calculate the noise spectrum from both sides of the FDT, 1) by computing the
time-correlation $\avg{V_M(0)V_M(t)}$ from the response function, then using \Eq{eqn:Somega};
2) by computing the frequency-dependent impedance $Z({\omega})$ of a spin valve, which consists of an
electrical and a magnetic contribution: $Z({\omega}) = R_E + Z_M({\omega})$, then making use of the
Johnson-Nyquist formula \Eq{eqn:JN}. The electric part $R_E$ gives rise to a white
Johnson-Nyquist noise of $S_E = 4 k_B T R_E$. In this section, we focus on method 1), and
calculate the voltage noise spectrum for two special cases with $\mm_0\|\hxx$ (perpendicular
case) and $\mm_0\|\hzz$ (parallel case). In Appendix \ref{sec:Zomega}, we reproduce the
spectrum for $\mm_0\|\hxx$ by using method 2).


In bulk ferromagnets, magnetic moment dissipation is parameterized by the Gilbert damping
constant ${\alpha}_0$, which is associated with thermal fluctuations of the direction of the
magnetization vector. \cite{brown_thermal_1963} The magnetization fluctuations are caused by a
fluctuating torque from the lattice, which is represented by a thermal random magnetic field
$\hh^0$: $-M_{tot}\mm{\times}\hh^0$. The auto-correlator of $\hh$ is \cite{brown_thermal_1963}
\begin{equation*}
	\avg{{\gamma}h_i^0(t){\gamma}h_j^0(0)} 
	= {2{\gamma}{\alpha}_0\kB T\over M_{tot}}{\delta}_{ij}{\delta}(t) = \Sigma_0{\delta}_{ij}{\delta}(t)
	\label{eqn:ht}
\end{equation*}
with $i, j= x, y$ (assuming that the easy axis is along $z$). Similarly, the ferromagnet loses
energy and angular momentum by spin pumping. The magnetic damping increment ${\alpha}'$ must be
accompanied by a fluctuating transverse spin current (torque) ${\bf I}_s^{fl}$ from the
contacts, \cite{foros_magnetization_2005} which can be represented by another random magnetic
field $\hh'$: ${\bf I}_s^{fl} = -M_{tot} \mm{\times}\hh'$ with auto-correlator
\cite{foros_magnetization_2005}
\begin{equation*}
	\avg{{\gamma}h_i'(t){\gamma}h_j'(0)} 
	= {2{\gamma}{\alpha}'\kB T\over M_{tot}}{\delta}_{ij}{\delta}(t) = \Sigma'{\delta}_{ij}{\delta}(t).
	\label{eqn:hp}
\end{equation*}
$\hh'$ and $\hh^0$ are statistically independent: $\avg{h_i'h_j^0} = 0$.

Including spin pumping from the magnetization fluctuations and the fluctuating spin current
from the contacts, the total instantaneous spin current through the F\big|N interface between
the spacer and the free layer is (see \Figure{fig:spinvalve}(b)):
\begin{equation}
	{\bf I}_s(t) = {\bf I}_s^{sp} + {\bf I}_s^{fl}
	= {M_{tot}\over {\gamma}}\smlb{{\alpha}'\mm{\times}\dmm - {\gamma}\mm{\times}\hh'}.
\label{eqn:Is}
\end{equation}

Due to the filtering by the static layer magnetization $\mm_0$, the spin current ${\bf I}_s(t)$
is converted into a charge current $I_c(t)$ with efficiency ${\eta}({\theta})$. If the imaginary part of
the mixing conductance is disregarded ($g_i = 0$ and ${\tau}_{op} = 0$), an electrical voltage
$V_M(t)$ is given by the same expression as \Eq{eqn:VP} with $\mm{\times}\dmm$ replaced by $\mm{\times}\dmm
- ({\gamma}/{\alpha}')\mm{\times}\hh'$:
\begin{equation}
	V_M(t) = R({\theta}){\eta}({\theta}){2e\over{\hbar}}\midb{\mm_0{\cdot}{\bf I}_s(t)}
	= W({\theta})~\mm_0{\cdot}{\bf f}(t)
	\label{eqn:VM}
\end{equation}
with $W({\theta}) = 2eR({\theta}){\eta}({\theta})eM_{tot}/{\gamma}{\hbar}$. Assuming that $\mm$ fluctuates around the
$\hzz$-axis ($\mm{\simeq}\hzz$), $\bf f$ reads (to the leading order in $\mm$ and $\hh'$)
\begin{subequations}
\begin{align}
	f_x(t) &= {\gamma}h'_y - {\alpha}'\dmmy, \\
	f_y(t) &= -{\gamma}h'_x + {\alpha}'\dmmx, \\
	f_z(t) &= {\gamma}(\mmy h'_x - \mmx h'_y) + {\alpha}'(\mmx\dmmy-\mmy\dmmx).
	\label{eqn:gxyz}
\end{align}
\end{subequations}

The spectrum $S_M$ depends on the direction of the polarizer. We need to compute, {\it e.g.}
$F_x(t) {\equiv} \avg{f_x(0)f_x(t)}$ when $\mm_0\|\hxx$ and $F_z(t) {\equiv} \avg{f_z(0)f_z(t)}$ when
$\mm_0\|\hzz$. The correlators of $\bf f$ are composed of those between $\dmm$ and/or $\hh'$,
which in turn can be expressed by the transverse magnetic susceptibility ${\chi}({\omega})$ (in frequency
domain) as the response to the magnetic field $\hh = \hh^0+\hh'+\hh''$ ($\hh'$ and $\hh''$
account for the random fields from the left and right interface of the free layer):
\begin{equation}
	\midb{\begin{matrix} m_x({\omega}) \\ m_y({\omega}) \end{matrix}}
	={\chi}({\omega})
	\midb{\begin{matrix} {\gamma}h_x({\omega}) \\ {\gamma}h_y({\omega}) \end{matrix}}.
	\label{eqn:mchih}
\end{equation}
All correlators can be calculated from \Eq{eqn:mchih}: \cite{smith_modeling_2001}
\begin{subequations} 
\label{eqn:mmmm}
\begin{align}
	\avg{m_i(t)m_j(0)} &= {\Sigma\over {\alpha}} {\int}{1\over {\omega}}{\chi}^-_{ij}({\omega})e^{-i{\omega}t}{d{\omega}\over2{\pi}},
	\label{eqn:mm} \\
	\avg{\dm_i(t)m_j(0)} &= -{\Sigma\over {\alpha}} {\int}i~{\chi}^-_{ij}({\omega})e^{-i{\omega}t}{d{\omega}\over2{\pi}},
	\label{eqn:dmm} \\
	\avg{\dm_i(t)\dm_j(0)} &= {\Sigma\over {\alpha}} {\int}{\omega} ~{\chi}^-_{ij}({\omega})e^{-i{\omega}t}{d{\omega}\over2{\pi}},
	\label{eqn:dmdm} 
\end{align}
\end{subequations}
with $\Sigma = \Sigma_0 + 2\Sigma'$ (the factor 2 comes from the two pumping interfaces) and
${\chi}^-_{ij}({\omega}) = [{\chi}_{ij}({\omega}) - {\chi}^*_{ji}({\omega})]/2i$, and
\begin{subequations} 
\label{eqn:mmhh}
\begin{align}
	\avg{m_i(t){\gamma}h'_j(0)} &= \Sigma' {\int}{\chi}_{ij}({\omega})e^{-i{\omega}t} {d{\omega}\over2{\pi}},
	\label{eqn:mh} \\
	\avg{\dm_i(t){\gamma}h'_j(0)} &= -\Sigma' {\int}i{\omega}~{\chi}_{ij}({\omega})e^{-i{\omega}t} {d{\omega}\over2{\pi}},
	\label{eqn:dmh}
\end{align}
\end{subequations}
By taking all correlators between $\mm, \dmm$, and $\hh'$ into account, we confirm that the DC
spin/charge current vanishes at thermal equilibrium: $\avg{{\bf I}_s(t)} = \avg{I_c(t)} = 0$ as
required by the second law of thermodynamics.

When $\mm_0\|\hxx$, the voltage noise power spectrum reads
\begin{equation}
	S_M^x({\omega}) = 2 W^2({\pi}/2) {\int}_{-{\infty}}^{\infty}dt e^{i{\omega}t} F_x(t),
	\label{eqn:SMx}
\end{equation}
Using \Eqss{eqn:mmmm}{eqn:mmhh}, we have
\begin{equation}
	F_x(t) = \Sigma'\midb{{\delta}(t) - {\alpha}' {\int}{\omega}~{\chi}^-_{yy}({\omega})e^{-i{\omega}t}{d{\omega}\over2{\pi}}}.
\label{eqn:Gxt}
\end{equation}
From \Eq{eqn:Gxt} and \Eq{eqn:SMx},
\begin{equation}
	S_M^x({\omega}) = 2 W^2({\pi}/2) \Sigma' \bigb{1- {\alpha}' {\omega}~\imm{{\chi}_{yy}({\omega})}},
	\label{eqn:SMx2}
\end{equation}
in terms of the imaginary part of the dynamic susceptibilities, {\it i.e.} the magnetic
dissipation. A measurement of the former therefore determines the latter, serving as an
alternative to e.g. FMR measurements.

When $\mm_0\|\hzz$, $S_M^z({\omega})$ follows from \Eq{eqn:SMx} by the replacement $F_x(t)\ra F_z(t)$
and $W({\pi}/2)\ra W(0)$. According to \Eq{eqn:gxyz}, this involves 4-point correlators, which can
be reduced to 2-point correlators by Wick's theorem: \cite{foros_resistance_2007,
triantafyllopoulos_central_2003} $\avg{abcd} = \avg{ab}\avg{cd} +\avg{ac}\avg{bd}
+\avg{ad}\avg{bc}$. After some tedious algebra, we reach
\begin{align}
\label{eqn:SMz2}
	&S_M^z({\omega}) = 2 W^2(0) \Sigma'^2 
	\left\{{1\over{\alpha}'}{\sum}_i \rem{{\chi}_{ii}(0)}\right.- \nn
	&\left. {\int}{d{\omega}'\over 2{\pi}} {{\omega}-2{\omega}'\over{\omega}'} {\sum}_{i, j}
	(-1)^{{\delta}_{ij}}{\chi}^-_{ij}({\omega}'){\chi}^-_{\bar{i}\bar{j}}({\omega}-{\omega}')\right\}
\end{align}
with $\bar{x} = y$ and $\bar{y} = x$. For the anti-parallel configuration ($\mm_0\|-\hzz$), the
formula is identical to \Eq{eqn:SMz2} except that $W(0)$ is replaced by $W({\pi})$. Note that ${\chi}$
for parallel and anti-parallel cases are different, as discussed in Appendix A.

With \Eq{eqn:SMx2} and \Eq{eqn:SMz2}, the calculation of the noise power spectrum reduces to
that of the magnetic susceptibility ${\chi}$ for the free layer magnetization. Similar to the
Gilbert damping for the free layer magnetization in a spin valve,
\cite{tserkovnyak_dynamic_2003} ${\chi}$ in general depends on the magnetization configuration of
the spin valve. We derive the angular dependent ${\chi}$ in Appendix \ref{sec:chi}.

For simplicity we continue with an isotropic form of the magnetic susceptibility for the free
layer magnetization ${\chi}$, which includes the effect of the spin pumping enhanced damping, but
not the multiple scattering of the spin pumping current within the spacer:
\begin{equation}
	{\chi}({\omega}) 
	= {1\over ({\omega}_0-i{\alpha}{\omega})^2 - {\omega}^2}
	\smlb{\begin{matrix}  {\omega}_0-i{\alpha}{\omega} & -i{\omega} \\ i{\omega} & {\omega}_0-i{\alpha}{\omega} \end{matrix}}
	\label{eqn:chi}
\end{equation}
with ${\alpha} = {\alpha}_0 + 2{\alpha}'$ and ${\omega}_0 = {\gamma}H_{\rm eff}$. Using this ${\chi}$, we find
\begin{align}
	S_M^x({\omega})
	&= 2 W^2({\pi}/2) \Sigma' \nn
	&\bigb{1-{\alpha}'{\alpha}{(1+{\alpha}^2){\omega}^4+{\omega}_0^2{\omega}^2\over [(1+{\alpha}^2){\omega}^2-{\omega}_0^2]^2+4{\alpha}^2{\omega}^2{\omega}_0^2}}.
	\label{eqn:SMx3}
\end{align}
A more accurate form of $S_M^x({\omega})$ (\Eq{eqn:SMx4}) as calculated in Appendix \ref{sec:Zomega}
using circuit theory is recovered by the method here by using the angle-dependent
susceptibility tensor \Eq{eqn:chi2} instead of \Eq{eqn:chi} in \Eq{eqn:SMx2}.

\begin{figure}[b]
	\includegraphics[width=1.0\columnwidth]{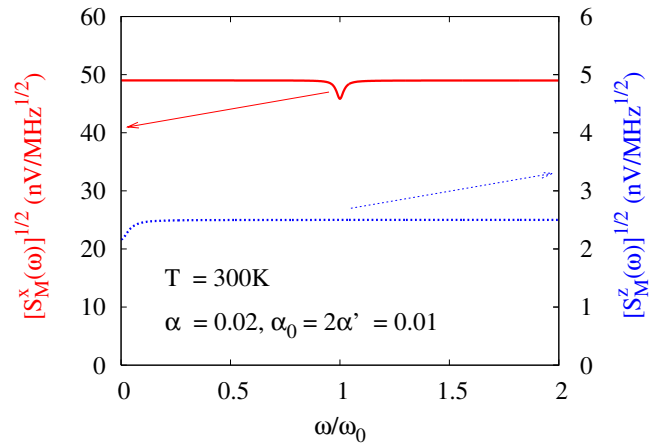}
	\caption{(Color online) $\sqrt{S_M^x({\omega})}$ (solid line with left scale) and
	$\sqrt{S_M^z({\omega})}$ (dashed line with right scale). }
	\label{fig:Vxz}
\end{figure}
The square root of $S_M^x({\omega})$ is plotted in \Figure{fig:Vxz} for the parameters in
\Table{tab:values}, where ${\eta}$ is replaced by its ballistic limit $p/2$ (left ordinate).
Assuming that the spin valve resistance $R$ is dominated by the interface resistances,
$R{\propto}1/A$, but does not depend on the free layer thickness $d$. Considering the volume ${\Omega}=Ad$
and ${\alpha}'{\propto}1/d$, $\Sigma'{\propto}{\alpha}'/{\Omega}{\propto}1/(Ad^2)$, the white noise background in $S_M^x({\omega})$
scales like $R^2{\Omega}^2\Sigma'{\propto}1/A$, and thus does not depend on $d$. The dip in
\Figure{fig:Vxz} at the FMR frequency is remarkable. Its depth is proportional to ${\alpha}'$ hence
inversely proportional to the free layer thickness, whereas its width is proportional to
${\alpha}{\omega}_0$.
The constant background of the spectrum is $\sqrt{S_M^x({\omega})}~{\simeq}~50$ nV/$\sqrt{\rm MHz}$,
whereas the dip is about 4 nV/$\sqrt{\rm MHz}$. For comparison, the root-mean-square of the
electrical contribution to the noise is $\sqrt{S_E({\omega})}=\sqrt{4\kB T R - S_M^x(0)}~{\simeq}~87$
nV/$\sqrt{\rm MHz}$.
\begin{table}
	\begin{tabular}{r|l|r}
		\hline
		Quantity	& Values	& Ref. \\
		\hline\hline
		$M_s$(Co)	& $1.42{\times}10^6$ A m$^{-1}$ & \onlinecite{katine_current-driven_2000}\\
		${\gamma}$(Co)& $1.9{\times}10^{11}$ (T s)$^{-1}$	& \onlinecite{bhagat_temperature_1974}\\
		${\alpha}_0$(Co)	& 0.01 					& \onlinecite{beaujour_magnetization_2006}\\
		$2{\alpha}'$(Co\big|Cu)& 0.01 					& \onlinecite{beaujour_magnetization_2006}\\
		\hline
		${\omega}_0$		& 10 GHz				& \\
		$p$			& 0.35					&
		\onlinecite{upadhyay_probing_1998,bass_current-perpendicular_1999}\\
		$R{\simeq}R_P$	& $0.57~{\Omega}$ 				& 
		Derived \footnote{$(R_{AP}-R_P)/R_{AP}~ {\approx} ~p^2$ and
		$R_{AP}-R_P = 0.073{\Omega}$ \cite{katine_current-driven_2000}, where P/AP stands
		for parallel/anti-parallel.} from Ref. \onlinecite{katine_current-driven_2000} \\
		$R_{\rm Sample}$& $1.6~{\Omega}$ 			& \onlinecite{katine_current-driven_2000} \\
		$T$			& 300 K 				& \\
		${\Omega} = A{\times}d$ & ($130{\cdot}130$ nm$^2$)${\times}2.5$ nm &
		\onlinecite{katine_current-driven_2000} \\
		\hline
	\end{tabular}
	\caption{Typical spin valve parameters (see text).}
	\label{tab:values}
\end{table}

When $\mm_0{\parallel}\hzz$, \Eq{eqn:SMz2} and \Eq{eqn:chi} lead to
\begin{equation}
	S_M^z({\omega})
	= 2 W^2(0){\Sigma'^2\over{\omega}_0}
	\midb{ {1\over {\alpha}'}- {\alpha} {(1+{\alpha}^2){\omega}^2+4{\omega}_0^2 \over (1+{\alpha}^2)^2{\omega}^2+4{\alpha}^2{\omega}_0^2}}.
	\label{eqn:SMz}
\end{equation}
The square root of $S_M^z({\omega})$ is plotted as the dashed curve in \Figure{fig:Vxz} (right
ordinate). In contrast to $S_M^x({\omega})$, the pre-factor in $S_M^z({\omega}) {\propto} R^2{\Omega}^2 {\Sigma'}^2/{\alpha}'
{\propto} 1/(A^2d)$, therefore the noise decreases with increasing $d$. This differs from the case
$\mm_0\|\hxx$ because the projection on the $\hzz$ axis involves the average deviation of $\mm$
from the equilibrium direction, which is inversely proportional to the volume. The divergence
at vanishing thickness is caused by the neglect of the finite transparency of very thin
magnetic layers for transverse spin currents. Similar to $S_M^x({\omega})$, the depth of the dip at
${\omega}= 0$ is proportional to ${\alpha}'$, hence inversely proportional to the layer thickness and has a
width proportional to ${\alpha}{\omega}_0$.

\section{Discussion} \label{sec:disc}

The spectrum $S_M^x({\omega})$ consists of three contributions, which can be seen from the
decomposition of $F_x(t) = \avg{f_x(t)f_x(0)} = F_x^{sp} + F_x^{fl} + F_x^{ab}$ with
\begin{subequations}
\begin{align}
	F_x^{sp}(t) &= {\alpha}'^2\avg{\dm_y(t)\dm_y(0)}, \\
	F_x^{fl}(t) &= {\gamma}^2\avg{h'_y(t)h'_y(0)}, \\
	F_x^{ab}(t) &= - {\alpha}'{\gamma} \midb{\avg{\dm_y(t)h'_y(0)} +\avg{\dm_y(0)h'_y(t)}}. 
\end{align}
\end{subequations}
These three contributions can be interpreted as: i) a spin pumping current $F_x^{sp}$, which
produces a peak at ${\omega}= {\omega}_0$, ii) a random torque (spin current) from the contact $F_x^{fl}$,
whose spectrum is white, and iii) the absorption of the random torque from the contact by the
magnetization $F_x^{ab} = -2 F_x^{sp}$, which gives a dip at ${\omega}= {\omega}_0$ with twice the
magnitude of i). The contacts therefore provide a white-noise random torque over the
ferromagnetic film, whereas the magnetization absorbs the noise power around ${\omega}_0$. The
spectrum $S_M^z({\omega})$ also consists of three contributions, but the absorption line of the
$\hzz$-component is centered at zero frequency because the fluctuations of the $\hzz$-component
magnetization do not have a characteristic frequency such as the $\hxx, \hyy$-components. In
inhomogeneous FM films, the single macrospin mode breaks up into different eigenmodes. The
noise power spectrum can then provide a ``fingerprint'' of the various eigenmode frequencies.

The three-point correlators arising in $\avg{V_M(0)V_M(t)}$ when $\mm_0$ is at arbitrary angles
in the $\hxx$-$\hzz$ plane vanish for normal distributions. The power spectrum is then a linear
combination of $S_M^x$ and $S_M^z$, depending on the angle with dips at both the FMR and zero
frequencies. The modeling of magnetic anisotropies by an easy axis is appropriate when the free
layer magnetization is oriented normal to the plane in axially symmetric pillars. In standard
pillars the dominant anisotropy is easy-plane, which leads to anisotropic fluctuations of the
magnetization. The results remain qualitatively similar, but become anisotropic in the
$\hxx$-$\hyy$ plane. For example, when a strong anisotropy constrains the fluctuations of $\mm$
to the $\hyy$-$\hzz$ plane, $S_M^y$ vanishes. We disregarded the imaginary part of the mixing
conductance in our calculation for the noise power spectrum. When it is included, the symmetric
dip in the power spectrum in \Figure{fig:Vxz} is skewed by $g_i$ similar to for the spin diode
effect discussed by Kupferschmidt {\it et al.} \cite{Kupferschmidt_theory_2006} and Kovalev
{\it et al.} \cite{kovalev_current-driven_2007}

For asymmetric spin valves, a non-monotonic angular dependence of the magnetoresistance and a
vanishing torkance at a non-collinear magnetization configuration has been demonstrated.
\cite{kovalev_spin_2002, manschot_nonmonotonic_2004, xiao_boltzmann_2004, boulle_shaped_2007,
rychkov_spin_2009} A sign change in torkance leads to a sign change in the charge pumping
voltage. The magnetic contribution to the thermal noise vanishes at the zero torkance point. 

\section{Summary} \label{sec:sum}

In conclusion, we find that a pumping voltage arises in a spin valve when the free layer
magnetization is in motion. The angular dependence of pumping voltage under FMR condition
provides detailed information of the spin transport in spin valves. The pumping voltage induced
by the thermal fluctuation of the free layer magnetization gives rise to additional voltage
noise, which is associated with the magnetization dissipation. Thus the equilibrium electronic
noise in a spin valve consists of two contributions: the Johnson-Nyquist noise associated with
the fluctuations of the charge, and magnetization related noise associated with the
fluctuations of the spins. The magnitude of these two contributions can be comparable. Unlike
the white Johnson-Nyquist noise, the latter is found to contain an absorption line at the FMR
frequency (and at zero frequency depending on the configuration) on top of an enhanced white
noise background. The noise spectrum can provide a fingerprint of the magnetic eigenmodes in
inhomogeneous structures.

\section*{Acknowledgment}

This work is supported by EC Contract IST-033749 ``DynaMax''. JX and GB thank SM for his
hospitality at Tohoku University. JX acknowledges support by H. J. Gao and S. Yi in Beijing and
X. F. Jin in Shanghai.

\appendix 

\section{Angular dependence of the transverse magnetic susceptibility in spin valves}
\label{sec:chi}

In this Appendix we calculate the transverse magnetic susceptibility for the free layer
magnetization in a spin valve when no external bias is applied ($I_c = 0$), {\it i.e.} the only
driving force is the thermal random field $\hh$.

For an isolated magnet, the dynamics are described by the Landau-Lifshitz-Gilbert equation:
\begin{equation}
	\dmm = -{\gamma}~\mm{\times}(\Heff + \hh) + {\alpha}_0~\mm{\times}\dmm ,
	\label{eqn:llg0}
\end{equation}
with the thermal magnetic field $\hh$ and bulk damping parameter ${\alpha}_0$. When we consider small
amplitude precession around the $\hzz$ direction (assuming $\Heff\|\hzz$), the Fourier
transform of the linearized LLG equation becomes
\begin{equation}
	\midb{\begin{matrix} m_x({\omega}) \\ m_y({\omega}) \end{matrix}}
	={\chi}_0({\omega})
	\midb{\begin{matrix} {\gamma}h_x({\omega}) \\ {\gamma}h_y({\omega}) \end{matrix}},
	\label{eqn:llg_w}
\end{equation}
with
\begin{equation}
	{\chi}_0({\omega})
	= {1\over ({\omega}_0-i{\alpha}_0{\omega})^2 - {\omega}^2}
	\smlb{\begin{matrix}  {\omega}_0-i{\alpha}_0{\omega} & -i{\omega} \\ i{\omega} & {\omega}_0-i{\alpha}_0{\omega} \end{matrix}}.
	\label{eqn:chi0}
\end{equation}

In spin valves, ${\chi}$ for the free layer magnetization depends on the relative orientation of
$\mm_0$ and $\mm$ because of the multiple scattering within the spacer, which depends on the
orientation of $\mm_0$, acts as spin-transfer torque on $\mm$. We now linearize \Eq{eqn:llg},
again assuming that $\mm$ fluctuates around the $\hzz$-axis with small amplitude
($\mm{\simeq}\hzz$):
\begin{subequations} 
\label{eqn:llg2}
\begin{align}
	\dm_x &= -{\omega}_0m_y - {\alpha}~\dm_y + {{\gamma}\over M_{tot}}N_{\rm st}^x + {\gamma}h_y, \\
	\dm_y &= +{\omega}_0m_x + {\alpha}~\dm_x + {{\gamma}\over M_{tot}}N_{\rm st}^y - {\gamma}h_x,
\end{align}
\end{subequations}
where ${\alpha} = {\alpha}_0 + 2{\alpha}'$ and $2{\alpha}'$ is the enhanced damping from the two interfaces of the free
layer.

The circuit theory \Eqss{eqn:Isp}{eqn:gsf} are coupled with the LLG equation in \Eq{eqn:llg2}
through $\dmm$ in \Eq{eqn:ctR} and $\II_R$ in \Eq{eqn:llg2} and they have to be solved
self-consistently. We assume that $\mm_0$ is static and tilted by an angle ${\theta}$ from $\hzz$,
{\it i.e. }$\hzz{\cdot}\mm_0 = \cos{\theta}$. \Eqss{eqn:Isp}{eqn:gsf} can be converted to scalar
equations by taking dot products with $\mm$, $\mm_0$, and $\mm_{\times} = \mm{\times}\mm_0$. Introducing
the projections of an arbitrary vector $\bf q$: $(q^0, q^m, q^{\times}) {\equiv} {\bf q}{\cdot}(\mm_0, \mm,
\mm_{\times})$, setting $g_{sf} = 0$ for simplicity, \Eqss{eqn:Isp}{eqn:gsf} become
\begin{subequations} 
\label{eqn:ct0mx}
\begin{align}
	0 &= I = {eg\over 2h}(2 {\mu}_L-p{\mu}_N^0) = -{eg\over 2h}(2 {\mu}_R-p{\mu}_N^m), \\
	I_L^0 &= {eg\over 2h}(2p {\mu}_L-{\mu}_N^0),\\
	I_L^m &= {eg\over 2h}(2p{\mu}_L-{\mu}_N^0)\cos{\theta}-{eg_r\over h}({\mu}_N^m-{\mu}_N^0\cos{\theta}), \\
	I_L^{\times} &= -{eg_r\over h}{\mu}_N^{\times}, \\
	I_R^0 &= -{eg\over 2h}(2p{\mu}_R-g{\mu}_N^m)\cos{\theta} \nn
	&+{eg_r\over h}({\mu}_N^0-{\mu}_N^m\cos{\theta}) +{eg_r\over 2{\pi}}\dm^{\times}, \\
	I_R^m &= -{eg\over 2h}(2p {\mu}_R-{\mu}_N^m), \\
	I_R^{\times} &= {e\over h}g_r{\mu}_N^{\times} -{eg_r\over 2{\pi}}\dm^0, \\
	0 &= I_L^0 - I_R^0 = I_L^m - I_R^m = I_L^{\times} - I_R^{\times},
\end{align}
\end{subequations}
where in the third and fifth equation above, we disregard the time-dependence of $m^0(t) =
\mm(t){\cdot}\mm_0$ because $\mm{\simeq}\hzz$ to leading order in the deviations. The solutions to
\Eq{eqn:ct0mx} are
\begin{equation} 
\label{eqn:IR0mx}
	I_R^0 = {\xi}_0 {eg_r\over 4{\pi}} \dm^{\times}, \quad
	I_R^m = {\xi}_m {eg_r\over 4{\pi}} \dm^{\times}, \quad
	I_R^{\times} = -{eg_r\over 4{\pi}} \dm^0,
\end{equation}
with
\begin{equation}
	{\xi}_0 = {1-{\nu} \over 1-{\nu}^2\cos^2{\theta}} \qand
	{\xi}_m = -{(1-{\nu}){\nu}\cos{\theta} \over 1-{\nu}^2\cos^2{\theta}}
\label{eqn:xi}
\end{equation}
and ${\nu} = [2g_r-g(1-p^2)]/[2g_r+g(1-p^2)]$. We now define the coordinate system in the plane
normal to $\hzz$:
\begin{equation}
	\hxx {\equiv} {\mm_0-\hzz\cos{\theta}\over\sin{\theta}}
	\qand \hyy {\equiv} \hzz{\times}\hxx {\approx} {\mm_{\times}\over\sin{\theta}}.
\label{eqn:xy}
\end{equation}
Therefore $\dm^0 {\approx} \dm_x\sin{\theta}$, $\dm^{\times} {\approx} \dm_y\sin{\theta}$, and
\begin{subequations}
\label{eqn:Nbfxy}
\begin{align}
	{{\gamma}\over M_{tot}}N_{\rm st}^x &{\approx} {{\gamma}{\hbar}\over 2eM_{tot}} {I_R^0-I_R^m\cos{\theta}\over\sin{\theta}} = \half {\xi}{\alpha}'\dm_y, \\
	{{\gamma}\over M_{tot}}N_{\rm st}^y &{\approx} {{\gamma}{\hbar}\over 2eM_{tot}} {I_R^{\times}\over\sin{\theta}} = -\half {\alpha}'\dm_x, 
\end{align}
\end{subequations}
with ${\xi} = {\xi}_0-{\xi}_m\cos{\theta}$, which is related to the angular dependent magnetic damping in Ref.
\onlinecite{tserkovnyak_dynamic_2003}.

Plugging \Eq{eqn:Nbfxy} into \Eq{eqn:llg2}, after linearization in terms of small fluctuations
about the $\hzz$-axis and Fourier transformation, we have
\begin{equation}
	\label{eqn:chii}
	\midb{\begin{matrix} {\gamma} h_x({\omega}) \\ {\gamma} h_y({\omega}) \end{matrix}} =
	\smlb{\begin{matrix} {\omega}_0 - i{\alpha}_y{\omega} & i{\omega} \\ -i{\omega} & {\omega}_0 - i{\alpha}_x{\omega} \end{matrix}} 
	\midb{\begin{matrix} m_x({\omega}) \\ m_y({\omega}) \end{matrix}},
\end{equation}
where ${\alpha}_x = {\alpha} - \half{\xi}{\alpha}'$ and ${\alpha}_y = {\alpha} - \half{\alpha}'$ reflect the damping anisotropy. From
\Eq{eqn:chii},
\begin{widetext}
\begin{equation}
	{\chi}({\omega}) 
	= \smlb{\begin{matrix} {\omega}_0 - i{\alpha}_y{\omega} & i{\omega} \\ -i{\omega} & {\omega}_0 - i{\alpha}_x{\omega} \end{matrix}}^{-1} 
	= {1\over (1+{\alpha}_x{\alpha}_y){\omega}^2-{\omega}_0^2+i({\alpha}_x+{\alpha}_y){\omega}_0{\omega}} 
	\smlb{\begin{matrix} {\omega}_0 - i{\alpha}_x{\omega} & -i{\omega} \\ i{\omega} & {\omega}_0 - i{\alpha}_y{\omega} \end{matrix}}.
	\label{eqn:chi2}
\end{equation}
\end{widetext}
Because ${\alpha}_y$ depends on angle ${\theta}$ through ${\xi}$, ${\chi}({\omega})$ also becomes angle dependent, {\it
i.e.} the magnetic susceptibility function for the free layer magnetization in a spin valve is
in general angular dependent, for the same reason as the magnetic damping in spin valves.
\cite{tserkovnyak_dynamic_2003} \Eq{eqn:chi2} reduces to \Eq{eqn:chi} when we identify
${\alpha}_x{\simeq}{\alpha}_y{\simeq}{\alpha}$ by ignoring the back-flow correction to the damping.

\section{Spin valve impedance $Z({\omega})$ for $\mm_0\|\hxx$} \label{sec:Zomega}

Here, we calculate the frequency dependence of the impedance of a spin valve by applying a
small AC current at frequency ${\omega}$: $I({\omega})$. We consider the perpendicular case here, {\it
i.e.} $\mm_0\|\hxx$ or $\cos{\theta} = 0$, so that the circuit theory equations equal \Eq{eqn:ct0mx}
except that we allow for a {\it non-vanishing} charge current $I{\neq} 0$. Since we are now
interested in the deterministic response, the thermal random fields $\hh$ may be ignored. We
can then solve \Eqs{eqn:llg2}{eqn:ct0mx} self-consistently in the frequency domain. We find
that the impedance of the spin valve $Z({\omega}) = [{\mu}_L({\omega}) - {\mu}_R({\omega})]/e I({\omega})$ consists of two
parts, an electric part $R_E$ and a magnetic part $Z_M({\omega})$: $Z({\omega}) = R_E + Z_M({\omega})$:
\begin{subequations}
\begin{align}
	R_E &= {4\over G_0}{1\over g} + {\eta}^2 R^2G_0 g(1-p^2) + {\eta}^2 R^2 G_0 g_r,\\
	Z_M^x &= {\eta}^2 R^2 G_0 g_r{\times}\nn
	&\midb{1- {{\alpha}'{\omega}({\alpha}_y{\omega}+i {\omega}_0)\over (1+{\alpha}_x{\alpha}_y){\omega}^2-{\omega}_0^2+i({\alpha}_x+{\alpha}_y){\omega}{\omega}_0}},
	\label{eqn:Z}
\end{align}
\end{subequations}
where ${\eta} = {\eta}({\pi}/2)$ and $R = R({\pi}/2)$ are the polarization factor and the DC resistance for
the spin valve at ${\theta} = {\pi}/2$ or $\mm_0\|\hxx$. $R_E$ consists of the resistances associated
with the electrical dissipation and half of the magnetic dissipation from the interface with
the static magnetization which does not emit a spin pumping current.

By the Johnson-Nyquist formula \Eq{eqn:JN}, the noise spectrum $S_M^x({\omega})$ is given by
\begin{widetext}
\begin{equation}
	S_M^x = 4k_B T~\rem{Z_M^x({\omega})}
	= 2 W^2({\pi}/2) \Sigma' 
	\bigb{1- {\alpha}'{{\alpha}_y(1+{\alpha}_x{\alpha}_y){\omega}^4+{\alpha}_x{\omega}^2{\omega}_0^2\over
	[(1+{\alpha}_x{\alpha}_y){\omega}^2-{\omega}_0]^2+({\alpha}_x+{\alpha}_y)^2{\omega}^2{\omega}_0^2}}. 
	\label{eqn:SMx4}
\end{equation}
\end{widetext}
This equation is identical to \Eq{eqn:SMx3} when we that the limit ${\alpha}_x{\simeq}{\alpha}_y{\simeq}{\alpha}$. This
difference comes from the approximate form of ${\chi}$ in \Eq{eqn:chi}. If we use \Eq{eqn:chi2},
then the $S_M^x$ calculated from \Eq{eqn:SMx2} will be exactly the same as \Eq{eqn:SMx4}.

The frequency dependent impedance for $\mm_0\|\hzz$ is second order in $m_{x,y}$, therefore it
is not so straightforward to calculate, which can also be seen from the non-trivial convolution
in \Eq{eqn:SMz2} calculated from magnetic susceptibility functions.


\end{document}